\documentclass[reprint,
amsmath,amssymb,
aps,
pra,
prl,
rmp,
prstab,
prstper,
floatfix,
]{revtex4-1}

\DeclareUnicodeCharacter{0308}{HERE!HERE!}
\DeclareUnicodeCharacter{0306}{HERE!HERE!}
\DeclareUnicodeCharacter{0090}{HERE!HERE!}
\usepackage{graphicx}
\usepackage{tabularx}
\usepackage{appendix}
\usepackage[utf8]{inputenc}
\usepackage{dcolumn}
\usepackage{bm}

\begin{document}

\preprint{APS/123-QED}

\title{Suppression of stacking order with doping in 1T-TaS$_{2-x}$Se$_{x}$}

\author{Sharon S. Philip}
\affiliation{University of Virginia, Charlottesville, VA 22904, USA}
\author{Despina Louca*}
\affiliation{University of Virginia, Charlottesville, VA 22904, USA}
\author{J. C. Neuefeind}
\affiliation{Oak Ridge National Laboratory, Oak Ridge, TN 37830, USA}
\author{Matthew B. Stone}
\affiliation{Oak Ridge National Laboratory, Oak Ridge, TN 37830, USA}
\author{A. I. Kolesnikov}
\affiliation{Oak Ridge National Laboratory, Oak Ridge, TN 37830, USA}



\begin{abstract}
In 1T-TaS$_{2-2x}$Se$_{2x}$, the charge density wave (CDW) state features a star of David lattice that expands across layers as the system becomes commensurate on cooling. The layers can also order along the c-axis and different stacking orders have been proposed. Using neutron scattering on powder samples, we compared the stacking order previously observed in 1T-TaS$_{2}$ as the system is doped with Se. While at low temperature, a 13c layer sequence stacking was observed in TaS$_{2}$, this type of ordering was not evident with doping. Doping with Se results in a nearly commensurate state with the Mott state suppressed which may be linked to the absence of the layer stacking.
\end{abstract}

\maketitle

\section{\label{sec:intro}Introduction}

Quasi-two dimensional (2D) in nature, transition metal dichalcogenides (TMDs) 1T-MX$_{2}$ (M = Ti, Ta and X = S, Se, Te) are prone to electronic instabilities \cite{wilson1975charge}. 1T-Ta(S/Se)$_{2}$ exhibits an incredibly rich phase diagram with multiple charge density wave (CDW) transitions emerging as a function of temperature and upon doping. In 1T-TaS$_{2-2x}$Se$_{2x}$, macroscopic behaviors such as CDW and superconductivity \cite{liu2013,liu2016nature,wang2020band,Sipos2008} have been observed, and more recently, a quantum spin liquid (QSL) has been proposed in 1T-TaS$_{2}$ as well \cite{law20171t}. In the typical Peierls model for CDW order \cite{rossnagel2011origin}, the instability of the coupled electron-lattice system brings a structural phase transition that is driven by strong electron-phonon coupling \cite{chen2023ambipolar,sun2018hidden,butler,zhang2023stacking}. The CDW formation can bring electron localization where displacements along phonon modes lower the total electronic energy by opening up a gap at the Fermi level, E$_{F}$ \cite{warawa2023combined,hansen2023collective}. This scenario, although applicable to simple one-dimensional systems, does not fully describe the case of 1T-TaS$_{2-x}$Se$_{x}$ where the CDW behavior is intertwined with the opening of a Mott gap \cite{europhysics}. The origin of the CDW has been highly debated in TMDs \cite{wegner2020,chatterjee2015emergence,van2010alternative}. The Fermi surface nesting scenario most often does not apply. Existing models of the CDW order are broadly classified into three types: in one, it involves an excitonic condensation mechanism; in two, it involves a Jahn-Teller-like distortion mechanism; and in three, it involves a hybrid model, a combination of Jahn-Teller and exciton condensation\cite{van2010alternative}.

1T-TaS$_{2}$ exhibits a strong CDW instability and electronic localization that leads to several interesting effects. Upon cooling from high temperatures, three main phases form: the high temperature incommensurate CDW (ICDW), the intermediate temperature nearly commensurate CDW (NCCDW) and the low temperature commensurate CDW (CCDW) \cite{Ang2015}. The ICDW appears below ~540 K on cooling from the high temperature metallic state, with a transition from the $P\overline{3}m1$ crystal symmetry shown in Fig. 1(a) to the $P\overline{3}$ structure shown in Fig. 1(b). This transition leads to displacements of Ta ions that gives rise to the well-known star of David motifs. Upon cooling from the normal, high temperature metallic state, systematic displacements of the transition metal Ta leads to a star of David formation consisting of 13 Ta ions, in-plane. Domains of these formations expand to a commensurate CDW phase on cooling. Distinct from other CDW systems, in 1T-TaS$_{2}$, the commensurate CDW state is accompanied by a metal-insulator (MI) transition that has been proposed to arise either from Mott localization or from disorder induced Anderson localization. Important to the MI behavior are the orbital ordering and out of plane correlations, as well as layer stacking order.

In the ICDW, the stars have limited ordering in-plane. Further cooling leads to the ICDW becoming NCCDW at T=350 K, where the $\sqrt{13}\cdot\sqrt{13}$ structural modulation first appears with a 12$^{o}$ tilt relative to the original ab-plane. An expansion of the star of David motifs occurs in-plane \cite{spijkerman1997x}. Below 180 K, the $\sqrt{13}\cdot\sqrt{13}$ structural modulation persists with a rotation of 13.9$^{o}$ relative to the plane while the CDW becomes commensurate. The steps in the CDW transitions coincide with the kinks observed in the transport \cite{fazekas1979electrical} as the system goes from the metallic to the insulating state. On the other end of the phase diagram, in 1T-TaSe$_{2}$, the CCDW sets in at T $\approx$ 430 K, and the system shows no MI transition. It remains metallic down to the lowest temperature according to  transport data \cite{PhysRevLett.32.882}. Between these two ends, superconductivity emerges upon doping that coexists with a broad NCCDW region in the phase diagram \cite{Ang2015}. The coexistence of superconductivity with CDW domains has been observed in other TMDs such as in the 2H polytype and in other systems such as the cuprates \cite{canfield,PhysRevLett.125.186401,egami2000,lee2014coexistence}. 

The electronic bands appear to undergo a continuous change with decreasing temperature in going though the many transition steps \cite{Smith1985,PhysRevB.69.245123,PhysRevLett.81.1058}. In the absence of high temperature angle resolved photoemission spectroscopy (ARPES) due to resolution, there is no apparent nesting of the Fermi surface and a CDW gap is not necessarily located at the $\Gamma$ point. Measurements suggested that the gap appears elsewhere in k-space \cite{rossnagel2011origin,Fei}. The domain-like CDW structures of the NCCDW and ICDW states in 1T-TaS$_{2}$ are discommensurate and semi-metallic, but when the CDW becomes commensurate, the Fermi surface disappears and either a Mott-Hubbard localization or a disorder induced Anderson localization sets in \cite{Fei}. Across the NCCDW-CCDW boundary, the Fermi surface is continuously reduced. This effect is convoluted by d-electron localization that opens up an energy gap. In the CCDW phase, the gap is fully present, leading to a semiconducting state with about a 200 meV bandgap \cite{rossnagel2011origin}. 

In the normal phase above 540 K, the Ta 5d band at the $\Gamma$ point should be above E$_{F}$ \cite{PhysRevB.69.125117}. As the system goes through the NCCDW phase, this band becomes visible with ARPES. Further cooling to the CCDW state, band folding is observed because of the smaller Brillouin zone between 180 and 160 K, and an abrupt energy shift occurs due to opening of the energy gap \cite{wang2020band}. The loss of the Fermi surface continues with further cooling while the CDW gap continues to grow. The first order transition seen in the transport at 180 K on cooling is most likely due to a Mott-Hubbard localization \cite{ritschel2015orbital,Ritschel2018}. On warming, a different behavior is observed where the resistivity exhibits a hysteresis, with its value dropping ~280 K, marking the CCDW-NCCDW transition. This has been attributed to be due to changes in the c-axis stacking order \cite{Lee2019,wang2020band}.

We report on the nature of the layer stacking order with temperature and doping. Earlier, we observed that the c-axis expands in the CCDW phase of 1T-TaS$_{2}$ on warming but drops at the crossover between the CCDW-NCCDW transition \cite{philip2023local}. It has been suggested that the localization of the d-electrons that brings the gap in the electronic structure depends on the expansion of the c-axis \cite{ritschel2018stacking,ritschel2015orbital}. This in turn is related to the c-axis stacking order where changes in the interlayer coupling might drive the Mott transition. Neutron diffraction measurements confirmed the presence of 13c stacking order that disappears on warming across the CCDW-NCDW transition in 1T-TaS$_{2}$. The appearance of the 13c layer sequence is expected to drive the Mott localization. The 13c stacking sequence was previously suggested in Ref. \cite{Scruby1975TheRO} from X-ray diffraction data down to 80 K. This study extends the data down to 2 K.
Moreover, from single crystal measurements, we previously identified a 3c layer stacking as well, that commences in the ICDW state and continues to grow through the NCCDW to CCDW crossover \cite{PhysRevB.107.184109} It is possible that both the 3c and 13c coexist at low temperatures in 1T-TaS$_{2}$. Our single crystal data only reached 150 K. With doping, the neutron diffraction data clearly indicate that the 13c structure is suppressed. Its signature diffraction peak around 0.6 \AA\ is not observed with doping. At the same time, it is not clear what happens to the 3c stacking with doping. Further experiments using single crystals are underway to elucidate the doping dependence of the 3c order.

\section{\label{sec:results}Results and Discussion}

Shown in Fig.1(a) is the hexagonal crystal structure of the high temperature undistorted lattice. Layers of the transition metal are separated by the chalcogen ion creating a quasi-2D lattice where weak interlayer interactions are expected due to the van der Waals nature of the forces holding the layers together. However, orbitals play an important role in this TMD and out of plane electron correlations lead to layer ordering and a gap in the density of states. The out-of-plane coupling is important to understand the electronic characteristics of these materials where band structure calculations suggested that opening a gap at the $\Gamma$ point depends on the orbital order and out-of-plane stacking \cite{ritschel2015orbital,ritschel2018stacking}. Also shown in Fig.1(a) is the low temperature crystal structure in the CCDW phase where the high temperature cell has undergone a $\sqrt{13}\cdot\sqrt{13}$ structural expansion and a rotation of 13.9$^{o}$ relative to the primary axis. The star formation is the result of Ta displacements towards the middle Ta ion. The extent to which the star lattice spreads in the ab-plane depends on temperature. The star clusters expand on cooling giving rise to large domains in the CCDW state that are highly ordered, but become disordered on warming, breaking up into domains with star formations separated by regions of undistorted lattice. Three samples were measured using neutron scattering and the diffraction data are shown in Figs. 1(c) and 1(d). Powder samples of TaS$_{2}$, TaSSe and TaSe$_{2}$ were measured as a function of temperature. At 300 K, several superlattice reflections are indicated that might be due to 3c stacking structure. Similarly, at 2 K, the same superlattice reflections are observed as indicated. However, the intensity of the peaks is too small to be discerned from the powder diffraction data and single crystal experiments will help elucidate their presence.  

The reciprocal lattice vector $\mathbf{Q}$ was calculated using $\mathbf{Q} = h\mathbf{a_0^{*}} + k\mathbf{b_0^{*}} + l\mathbf{c^{*}} + m_1\mathbf{q^{1}} + m_2\mathbf{q^{2}}$, following the formalism introduced in~\cite{spijkerman1997x}. The modulation wave vectors $\mathbf{q^{1}} = \sigma_1\mathbf{a_0^{*}} + \sigma_2\mathbf{b_0^{*}}$ and $\mathbf{q^{2}} = -\sigma_2\mathbf{a_0^{*}} + (\sigma_1 + \sigma_2)\mathbf{b_0^{*}}$ for the CCDW phase were obtained based on the commensurate wave vector parameters $\sigma_1$ and $\sigma_2$. The reciprocal lattice commensurate wave vector can be described as $\mathbf{q}_{ccdw} = (\sigma_1\mathbf{a_0^{*}} +\sigma_2\mathbf{b_0^{*}})$ and for $\sqrt{13}\cdot\sqrt{13}$ in-plane translation, $\sigma_1$ and $\sigma_2$ values are 0.2308 and 0.0769 respectively. In the NCCDW phase, $\sigma_1$ and $\sigma_2$ values are 0.2448 and 0.0681 respectively. Shown in Fig. 1(e) is a plot of the diffraction pattern at very small momentum transfers, Q. At 5 K, a superlattice reflection belonging to 13c ordering is observed in 1T-TaS$_{2}$ at low temperatures. This reflection only appears from the 13c stacking order and not from the 3c order, even though most of the higher order reflections overlap between the two stacking models as shown in Fig. 1(f). The calculated positions of the satellite peaks corresponding to $3\ c$ stacking order ($\mathbf{c^*} = \mathbf{c_0^*}/3$) and $13\ c$ stacking order ($\mathbf{c^*} = \mathbf{c_0^*}/13$) are shown. Also shown in Fig. 1(e) are data for TaSSe in the same region of momentum transfer. In the TaSSe data, the reflection $\sim$ 0.6 \AA$^{-1}$ is notably absent which indicates that there is no 13c ordering in the superconducting state. A similar measurement was carried out for 1T-TaSe$_{2}$ and even though the sample was not a single phase of 1T, no evidence for the 0.6 \AA$^{-1}$ was observed. This indicates that stacking order might only be present in 1T-TaS$_{2}$.


\begin{figure}[h!]
\includegraphics[width=8.3 cm]{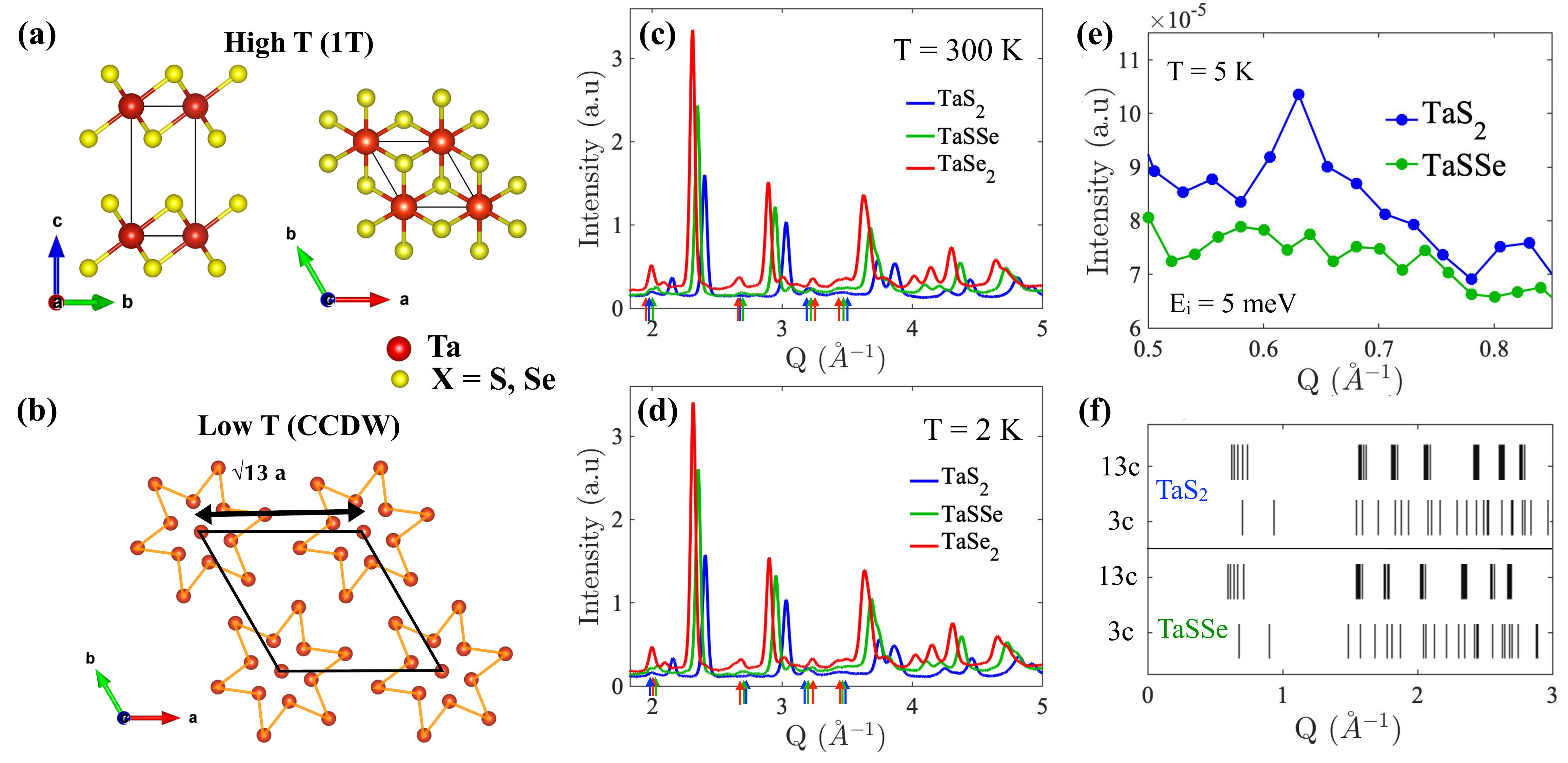}
\caption{(\textbf{a}) The high-temperature and low temperature crystal structure of 1T-TaX$_{2}$. The lattice symmetry is the trigonal $P\overline{3}m1$ at high temperature which becomes $P\overline{3}$ at low temperatures. The star is the result of the Ta displacements. (\textbf{b}) A plot of the diffraction pattern at low temperatures showing the presence of the 13c superlattice in 1T-TaS$_{2}$ although absent in 1T-TaSSe. The calculated peak positions of 1T-TaSSe corresponding to 3c and 13c stacking order is shown at the bottom. (\textbf{c}) The neutron powder diffraction data collected at 300 K compared among 1T-TaS$_{2}$, 1T-TaSSe and 1T-TaSe$_{2}$. All data are fit well using the $P\overline{3}$ symmetry. The diffraction peaks shift to the left with doping because Se is nominally a larger ion than S. (\textbf{d}) The neutron powder diffraction data collected at 2 K are shown. The arrows mark the positions of the CDW superlattice reflections.(\textbf{e}) The diffraction data plotted at very low Q indicate a superlattice peak corresponding to the 13c stacking order present in 1T-TaS$_{2}$. (\textbf{f}) A plot of the expected positions of 13c and 3c stacking order.}
\label{fig:structure}
\end{figure}  

\begin{figure}[h!]
\includegraphics[width = 8 cm]{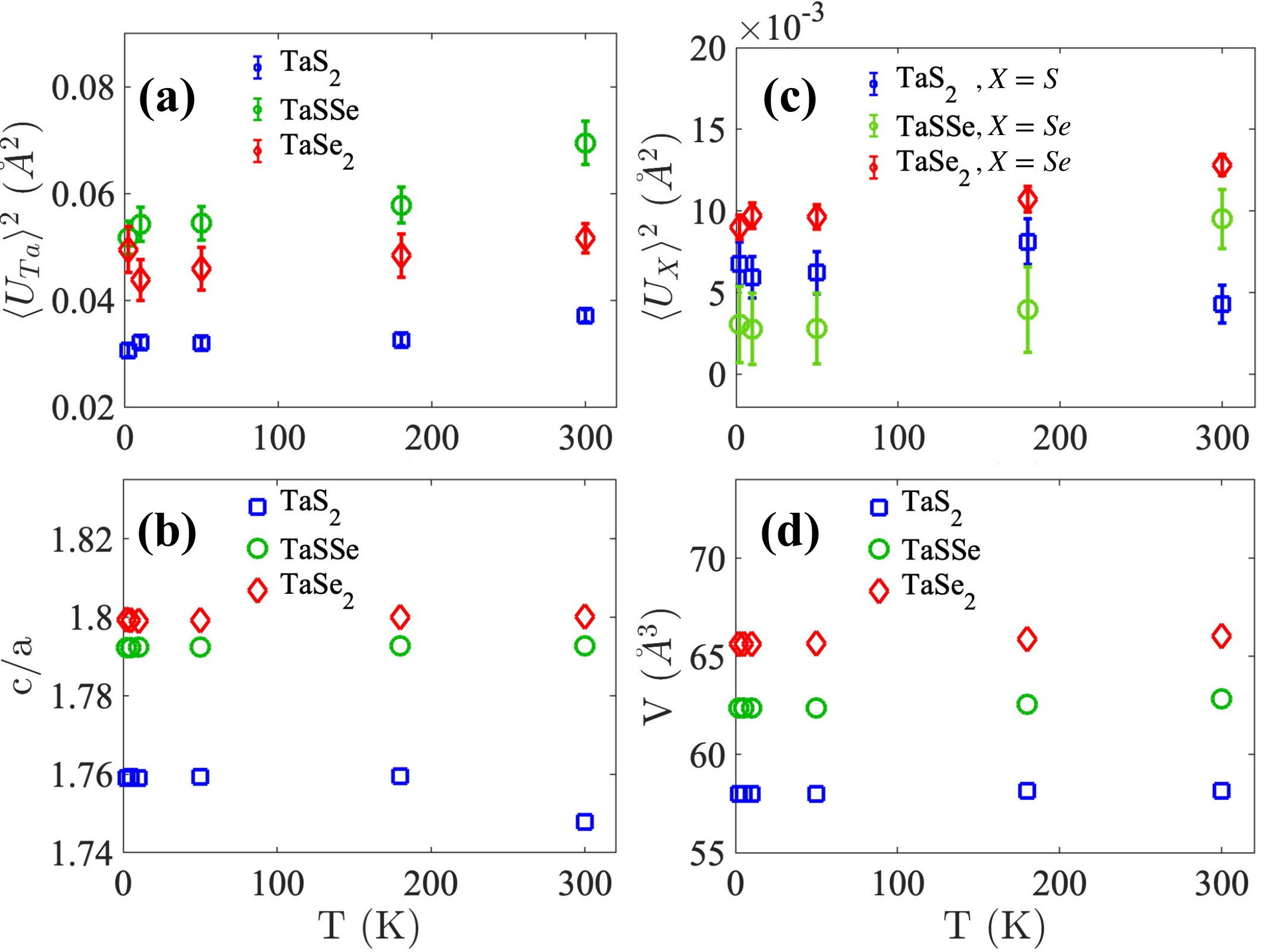} 
\caption[width 
 = \textwidth]{(\textbf{a}) The Ta atomic displacement <U>$^2$ is shown as a function of temperature for the three compositions. Of the three, superconducting TaSSe shows the largest thermal factors. Shown in (\textbf{c}) are the S and Se thermal factors for the three compositions. In (\textbf{b}) is a plot of the c/a ratio for the three compositions and in (\textbf{d}) is a plot of the unit cell volume. The c/a ratio in 1T-TaS$_{2}$ shows a decline on warming past 200K. }
\label{fig:thermalu}
\end{figure}

The temperature and composition dependence of the Ta and S/Se thermal factors, $\langle U \rangle^{2}$, lattice constants and unit cell volume are plotted in Fig. ~\ref{fig:thermalu}. As a function of composition, superconducting TaSSe has the largest thermal factor for the Ta ion that continues to increase on warming. Shown in Fig. ~\ref{fig:thermalu}(c) are the thermal factors for S and Se. Fig. ~\ref{fig:thermalu}(b) is a plot of the c/a ratio. In the case of TaSSe and TaSe$_{2}$, the c/a ratio is almost constant as a function of temperature which indicates that the unit cell expands uniformly in the a- and c-direction. However, in TaS$_{2}$, the ratio drops between 200 and 300 K because of the contraction of the c-lattice constant as previously observed in our earlier study \cite{philip2023local} and by others \cite{petkov2022atomic}. The contraction of the c-axis corresponds to the transition from the commensurate to the nearly commensurate state. We observed that this transition is coupled to the disappearance of the 13c ordering. Shown in Fig. ~\ref{fig:thermalu}(d) is the unit cell volume for the three compositions as a function of temperature. 

\begin{figure}[h!]
\includegraphics[width = 8 cm]{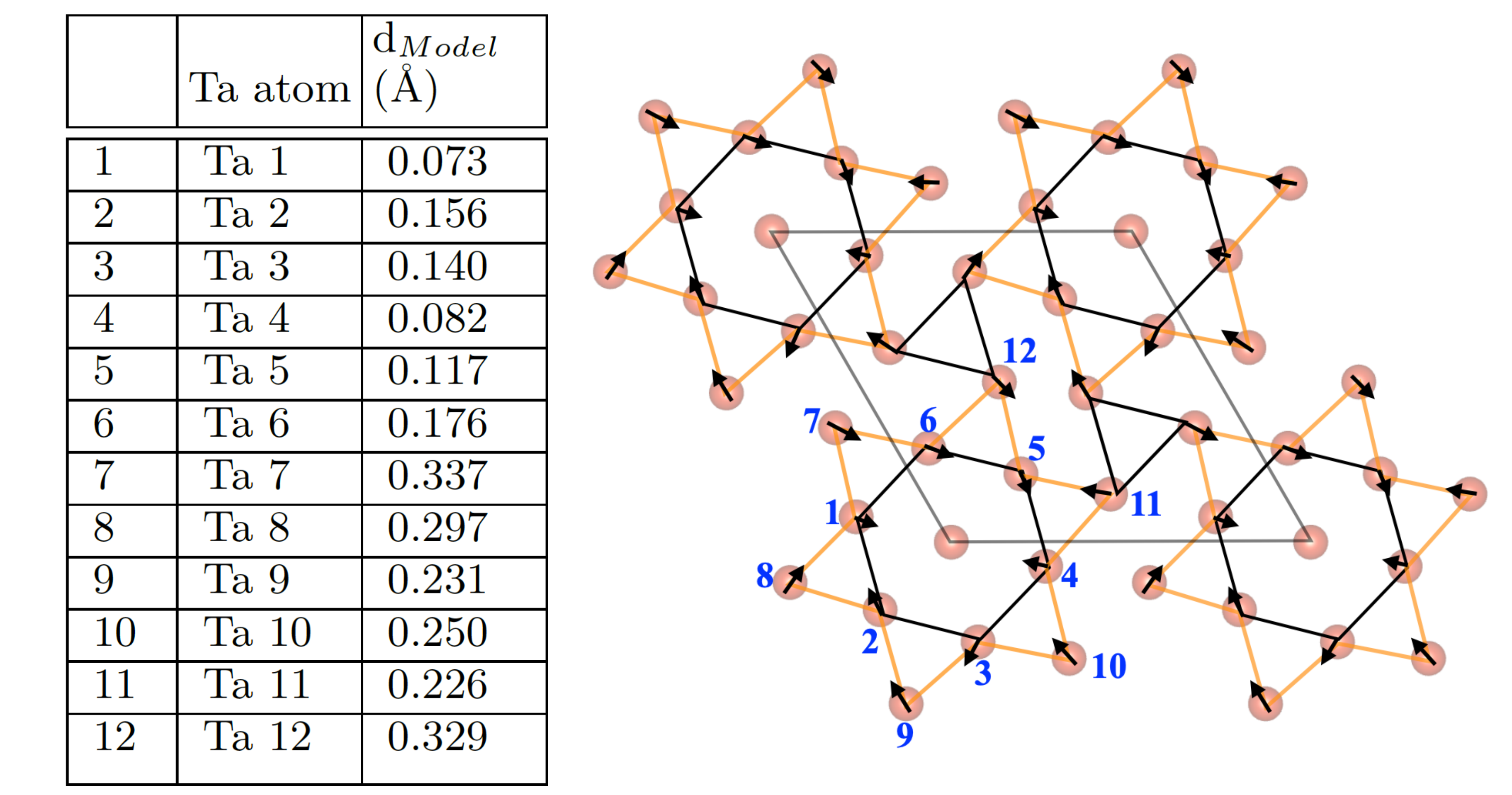} 
\caption[width 
 = \textwidth]{(\textbf{a}) A list of the Ta distortions obtained from fitting the local atomic structure. The Ta atoms make up the star of David. The  center Ta ion does not move by symmetry.}
\label{fig:table}
\end{figure}

Fig.~\ref{fig:table} shows the results from the pair density function (PDF) analysis of the Ta displacements from the local structure at 2 K. The local structure is obtained by Fourier transforming the diffraction data shown in Fig. 1, to obtain the pair correlation function, G(r). Fitting of the G(r) with a local model results in the distortions shown in the table of Fig.~\ref{fig:table}(a). Local Ta distortions are listed for the 12 Ta ions shown in ~\ref{fig:table}(b). The 13th center Ta ion does not move by symmetry. This indicates that even after the transition from the $P\overline{3}m1$ to the $P\overline{3}$ symmetry, locally the stars are distorted due to displacements of Ta in the directions shown with the arrows in the star lattice on the right. Moreover, the Ta ions are not all displaced in a symmetric way. This implies that the local trigonal symmetry is broken but that there is, nonetheless, long-range order of the star of David motifs in-plane. Similar distortions were observed in all three compositions with the results listed for 1T-TaS$_{2}$. 

\section{\label{sec:mat}Materials and Methods}

Powders were prepared using solid-state reaction. The neutron powder diffraction measurements were performed to investigate the structure through the multiple CDW steps. The time-of-flight (TOF) neutron measurements were carried out at the Nanoscale Ordered Materials Diffractometer (NOMAD/BL-1B) and at SEQUOIA (BL-17), a direct geometry spectrometer, at the Spallation Neutron Source (SNS) of Oak Ridge National Laboratory (ORNL) at temperatures ranging from 1.8 to 500 K. The aluminium can was used for SEQUOIA measurements and the empty can data were subtracted from the data. The reason SEQUOIA was used is that it reaches very small momentum transfers, not accessible to NOMAD. The diffraction data from NOMAD were analyzed using the Rietveld refinement to obtain the unit cell parameters characterizing the crystal structure \cite{toby2001expgui}, resulting in what is referred to as the average model. The pair density function (PDF) analysis~\cite{proffen2003structural,egami2003underneath} provides information on the local arrangement of atoms in real space without the assumption of periodicity. It was performed on the same neutron diffraction data as the ones used for the Rietveld refinement. NOMAD is a diffractometer with a large bandwidth of momentum transfer $Q$, and it provides the total structure function $S(Q)$. The $S(Q)$ was Fourier transformed into real-space to obtain the $G(r)$~\cite{warren1990x,pdfgetnpeterson2000}. The instrument background and empty sample container were subtracted from the $S(Q)$ and the data were normalized by a vanadium rod. A maximum $Q$ of 40 Å\textsuperscript{-1} was used. 

\section{Conclusions}

The formation of hetero Layer stacking can be engineered to enable new behaviors and new properties. For instance it has been theoretically proposed that stacking of the honeycomb ferromagnet CrI$_{3}$ has the potential to give rise to ferroelectricity \cite{ji2023general}. Moreover, stacking in moire superlattices can create polar domains because of local spontaneous polarization \cite{bennett2022electrically}. Hexagonal boron nitride was shown to exhibit ferroelectric switching in bilayers, leading to new concepts for functional heterostructures \cite{yasuda2021stacking}. Similarly, ferromagnetic heterostructures were demonstrated by stacking non-magnetic WS$_{2}$ with antiferromagnetic FePS$_{3}$. At the interface, the FePS$_{3}$ shows ferromagnetism \cite{gong2023ferromagnetism}.

Layer stacking in homostructure TMDs maybe similalry linked to the transport behavior. 1T-TaS$_{2}$ has an insulating CDW in contrast to other CDW dichalcogenides that have metallic CDW's. The reason for this is linked to Mott-Hubbard electron-electron correlations. Every star of David contributes one $\textit{5d}$ electron to a half filled narrow conduction band. In the 13c layer stacking, there is an odd number of electrons and in the presence of large Coulomb repulsion acting on the layers, the Mott-Hubbard transition occurs \cite{Fei}.
Density functional theory (DFT) calculations \cite{Lee2019} have shown that the insulating phase and the MI transition originate not from the 2D order of the stars of David but by the vertical order. Our results confirm the significance of interlayer coupling and the insulating property. Interlayer stacking order in the CCDW phase has been verified to be a 13c repeat unit cell. This result contradicts the notion that the stacking is partially disordered in the CCDW state. Hence it is less likely that Anderson localization drives the MI transition. This result also contradicts the bilayer stacking model.

\section{Acknowledgements}
A portion of this research used resources at the Spallation Neutron Source, a DOE Office of Science User Facility operated by Oak Ridge National Laboratory. We thank Dr. John Schneeloch (University of Virginia) for valuable inputs on the sample growth and Dr. Utpal Chatterjee for valuable discussions about the ARPES data.

\bibliography{TaSe2}
\end{document}